\documentclass[a4paper]{article}
\usepackage{graphicx,amsmath,amssymb}
\newcommand{\haak}[1]{\left(#1\right)}
\newcommand{\rhaak}[1]{\left [#1\right]}

\newcommand{\ahaak}[1]{\left\{#1\right\}}

\newcommand{\haakl}[1]{\left(#1\right.}
\newcommand{\haakr}[1]{\left.#1\right)}

\newcommand{\lhaakr}[1]{\left.#1\right |}

\newcommand{\half}{\frac{1}{2}}

\renewcommand{\imath}{{\rm i}}

\title{Osculating Random Walks on Cylinders} 
\author{Saibal Mitra and Bernard Nienhuis\\ 
Instituut voor Theoretische Fysica,\\ Universiteit van
Amsterdam,\\ 1018 XE Amsterdam,\\ The Netherlands\\
\texttt{saibalm@science.uva.nl, nienhuis@science.uva.nl}
}
\begin{document}
\maketitle
\begin{abstract}
We consider random paths on a square lattice which take a left or a
right turn at every vertex. The possible turns are taken with equal
probability, except at a vertex which has been visited before.
In such case the vertex is left via the unused edge. When the initial
edge is reached the path is considered completed. We also consider
families of such paths which together cover every edge of the lattice
once and visit every vertex twice. Because these paths may touch but not
intersect each other and themselves, we call them osculating walks.
The ensemble of such families is also known as the dense O$(n=1)$ model. 
We consider in particular such paths in a cylindrical geometry, with the
cylindrical axis parallel with one of the lattice directions. We
formulate a conjecture for the probability that a face of the lattice is
surrounded by $m$ distinct osculating paths. For even system sizes we
give a conjecture for 
the probability that a path winds round the cylinder. For odd system
sizes we conjecture the probability that a point is visited by a path
spanning the infinite length of the cylinder.
Finally we conjecture an expression for the asymptotics of a binomial determinant
\end{abstract} 

\section{Introduction}
In this article we present conjectures on the probability distribution
of random paths on an $L\times\infty$ square lattice with periodic
boundary conditions. The geometry of the lattice is thus a cylinder with
circumference $L$. 
When $L$ is even (assumed tacitly) the paths are closed loops with
probability one. When $L$ is odd (always stated explicitly) there is an
infinite path spanning the length of the cylinder.

The paths are allowed to have vertices in common but can not intersect. 
At each vertex the paths must turn, either to the left or to the right. 
These two possibilities are assigned equal probability, except when the
vertex has been visited before. In that case the path leaves the vertex
via the unused edge. This rule prevents (self) intersection of
paths. When a path reaches its initial edge it is considered
completed. We call paths of this type osculating walks. Osculating
lattice paths were considered  in \cite{brak}. 

\begin{figure}[t]
\begin{center}
\includegraphics[width=0.21\textwidth]{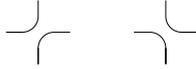}
\caption{The two vertices of the dense O(1) loop model}\vskip-5mm\label{fig:vrt}
\end{center}
\end{figure}

The conjectures are
obtained from the ground state of the dense O$(1)$ loop model. The dense
O$(n)$ 
loop model \cite{loop} can be defined as follows. The states of this
model are graphs consisting of paths of osculating walkers which together cover
all the edges of the lattice. The vertices can thus be in two states as
shown in fig.\ \ref{fig:vrt}. 
To a state consisting of $l$ closed paths (loops) a weight of $n^{l}$ is
assigned. We consider exclusively the case $n=1$, in which all
configurations have equal weight. 
The probability that a vertex is in a particular state is thus $\half$,
irrespective of the states the other vertices are in. Therefore, any
problem involving any finite number of osculating walkers can always be
reformulated as a computation of a probability in the O$(1)$ loop
model. In particular the probability distribution for the path of a
single osculating walker are precisely that of finding precisely that
path in the O$(1)$ loop model.

\begin{figure}[ht]
\begin{center}
\includegraphics[width=0.15\textwidth]{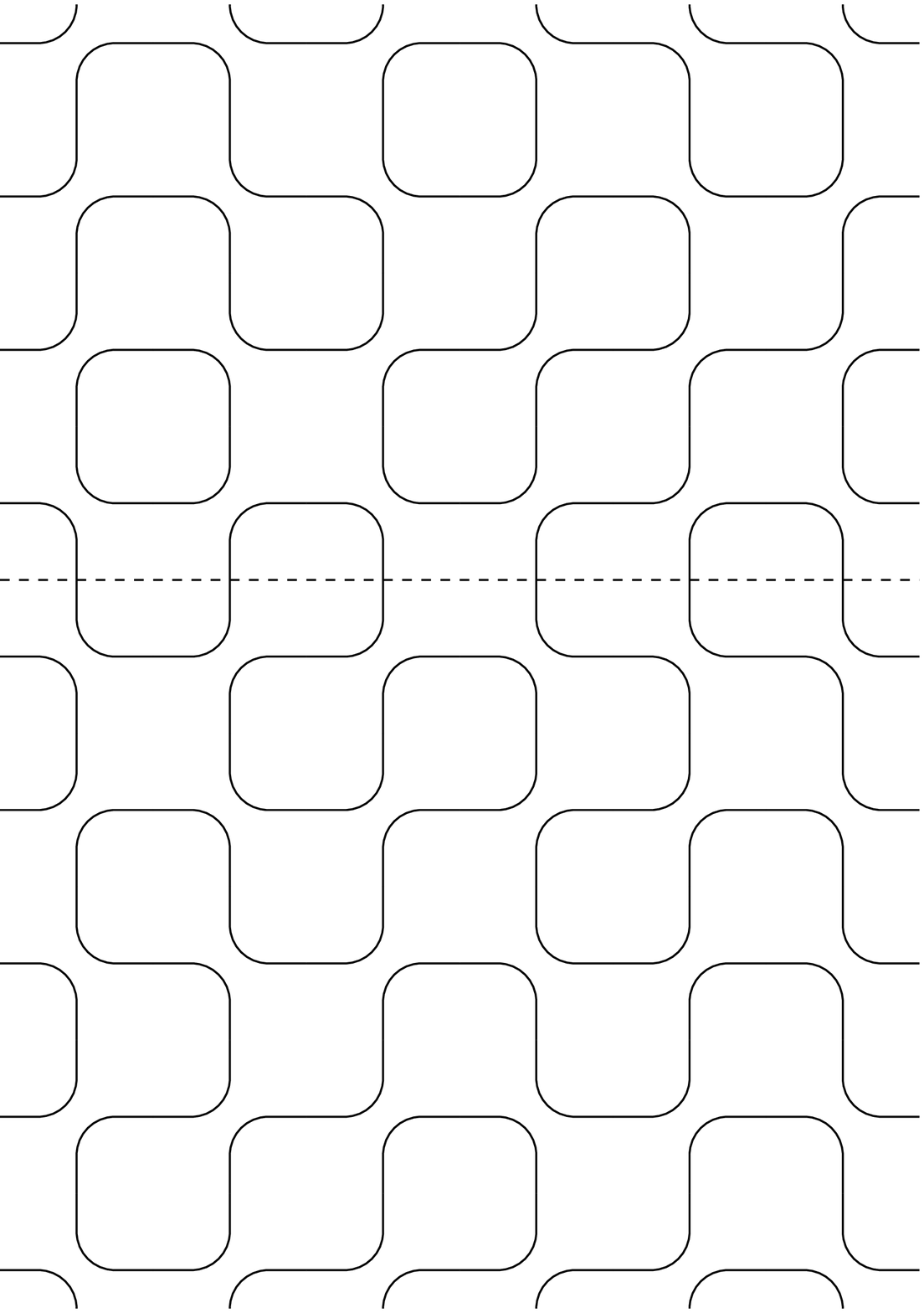}
\caption{Part of a typical configuration of the O$(1)$ loop model on a
$6\times \infty$ cylinder, periodic in the horizontal direction and
extending to infinity in both vertical directions.
The connectivity state of the half cylinder below the row indicated by the dashed line is $\haak{}^3$,
above the row it is $\haakr{}\haak{}\haakl{}\haak{}$.}\label{fig:O1}
\end{center}
\end{figure}

Recently, a number of conjectures about correlations of this model 
have been obtained. These involve probabilities that loops intersect 
a horizontal cut between vertices (henceforth referred to as a row) 
in a certain prescribed way. A row intersects loops at a total of $L$ points. 
We define the connectivity state of the row as the way these $L$ points are 
connected to each other by the loops via the half cylinder below the row. 
See fig.\ \ref{fig:O1} for an example. \vskip-4mm
 \vskip-5mm

Following \cite{mtr} we represent connectivity states by a string of parentheses. If a point at position $i$
is connected to a point at position $j$, then this is represented by a parenthesis at position $i$ matching with a parenthesis at position $j$. Because of periodic boundary conditions a left parenthesis doesn't have to be at the left of the matching right parenthesis.
The expression $\haak{\ldots}_{k}$ shall stand for $\haak{\haak{\haak{\ldots}}}$, where $k$ delimiters have been 
opened and closed, and the dots symbolize an arbitrary well nested configuration.
For instance $\haak{\haak{\haak{}}}$ will be denoted as $\haak{}_{3}$, and
$\haak{\haak{\haak{}\haak{\haak{}}}}$ will be denoted as $\haak{\haak{}\haak{}_{2}}_{2}$.
We will omit subscripts equal to~1. 
With a superscript we will denote a repeated concatenation of a structure with itself.
For instance $\haak{}_{2}^{k}$ stands for a sequence of $k$ $\haak{}_{2}$: 
$\haak{}_{2}\haak{}_{2}\ldots$.

A quantity of interest is the probability distribution over the set of all connectivity states. We denote the probability of finding a row in connectivity state $c$ as $P\rhaak{c}$. Denoting the conditional probability that a row is in state $c_{1}$ if the row below is in state $c_{2}$ as $P\rhaak{\lhaakr{c_{1}}c_{2}}$, we can write:
\begin{equation}\label{eval}
\sum_{c_{2}}P\rhaak{\lhaakr{c_{1}}c_{2}}P\rhaak{c_{2}}=P\rhaak{c_{1}}
\end{equation}

For finite $L$, the conditional probabilities $P\rhaak{\lhaakr{c_{1}}c_{2}}$ can be constructed by counting how many times
state $c_{1}$ can be obtained from $c_{2}$, by continuing the lines to
the next row in all possible ways. According to \eqref{eval} the
probability distribution $P\rhaak{c}$ is the right eigenvector of the
matrix $P\rhaak{\lhaakr{c_{1}}c_{2}}$. For $L$ not too large
$P\rhaak{c}$ is thus easily obtained. Computations of $P\rhaak{c}$ for
$L$ up to 18 have led to several conjectures
\cite{BdGN1,Strog2,Strog3,degier02112852, mtr}. 

A recent summary of these results is given in
\cite{degier02112852}. Here we review a few of these conjectures:
(i) The connectivity with the smallest probability is $\haak{}_{L/2}$. (ii)
The probability of any other connectivity is an integer multiple of the
probability of $\haak{}_{L/2}$. 
The reciprocals of $P\rhaak{\haak{}_{L/2}}$ for $L=2,4,6,8,10,12\ldots$
are: 2, 10, 140, 5544, 622908, 198846076,... 
This is a well known integer sequence. It is given by the number of
$L\times L$ alternating sign matrices invariant under a half turn
\cite{rob}. It is thus conjectured that (iii):
\begin{equation}
P\rhaak{\haak{}_{L/2}}=\rhaak{A_{\text{HT}}\haak{L}}^{-1}
\end{equation}
Here $A_{\text{HT}}\haak{L}$ is given by \cite{kup2}:
\begin{equation}
A_{\text{HT}}\haak{L}=2\prod_{k=1}^{L/2-1}\frac{3\haak{3k + 2}!\haak{3k -
1}!k!\haak{k
-1}!}{4\haak{2k + 1}!^2\haak{2k - 1}!^2}
\end{equation}
(iv) The event $\haak{}^{L/2}$ has the largest probability. The ratios $P\rhaak{\haak{}^{L/2}}\!/P\rhaak{\haak{}_{L/2}}$ for
$L=2,4,6,8,10,12\ldots$ are: 1, 3, 25, 588, 39204, 7422987,... 
This integer sequence is given by the number of
$\haak{L-1}\times\haak{L-1}$ alternating sign matrices invariant under
a half turn \cite{rob}, so that (v):
\begin{equation}
P\rhaak{\haak{}^{L/2}}=\frac{A_{\text{HT}}\haak{L-1}}{A_{\text{HT}}\haak{L}}
\end{equation}
The conjectured expression for $A_{\text{HT}}\haak{L-1}$ (note that $L-1$ is
odd) is given by \cite{rob}:
\begin{equation}
A_{\text{HT}}\haak{L-1} = \prod_{j=1}^{L/2-1}\frac{4}{3}\frac{\haak{3j}!^2
j!^2}{\haak{2j}!^4}
\end{equation}
Many more conjectures concerning this probability distribution have been obtained, also for other boundary conditions.
\section{Bulk Connectivity States}
A connectivity state, as defined in the previous section, specifies how points are connected to each other below a certain row. This concerns only configurations on the half cylinder. A convenient way to compute probabilities at a row on the complete cylinder is to consider separately the connectivity states on both half spaces below and above the row. We define the bulk connectivity at a certain row as the pair $\ahaak{c_{1},c_{2}}$,
where $c_{1}$ $\haak{c_{2}}$ is the connectivity state below (above) the row. The fact that all loop configurations are equally likely in the O$(1)$ loop model implies that the probability $P_{\text{b}}$ of having the bulk connectivity $\ahaak{c_{1},c_{2}}$ factorizes:
\begin{equation}
P_{\text{b}}\rhaak{\ahaak{c_{1},c_{2}}}=P\rhaak{c_{1}}P\rhaak{c_{2}}
\end{equation} 
The probability that a row is intersected by a loop in a certain way directly translates to the probability that a single osculating walker intersects a row in that way. We have obtained the following conjectures concerning single osculating walkers:
\begin{itemize}
\item The probability that an osculating walker will visit the left neighbor of the starting point is:
\begin{equation}
\frac{11L^{2}+4}{16\haak{L^{2}-1}}
\end{equation}
\item The probability that the path of an osculating walker winds round the cylinder is given as:
\begin{equation}\label{wnd}
\frac{A\haak{L}}{A_{\text{HT}}\haak{L}^{2}}
\end{equation}
where $A\haak{L}$ is the number of $L\times L$ alternating sign matrices \cite{kup2,zeil}, given by:
\begin{equation}
A\haak{L} =\prod_{j=0}^{L-1}\frac{\haak{3j + 1}!}{\haak{L + j}!}
\end{equation}
\end{itemize}
For large $L$ this last probability decays as $L^{-1/4}$.

So far we have only considered cylinders with even circumferences
$L$. If $L$ is odd, one has to modify the definition of the connectivity
states. On the top row of a half infinite cylinder there will be one
point that is not connected to any other point on the row. This point
will thus be on an ''open'' loop spanning the length of the
cylinder. This implies that no loop can wind round the cylinder, because
any such loop must necessarily cross the open loop, an event not allowed
in the O$(1)$ loop model. Interestingly, we conjecture that the
probability that a point on an odd $L$ cylinder is visited by this open
loop is given by the 
same formula \eqref{wnd} which for even $L$ gives the probability that a
point is on a loop spanning the circumference of the
cylinder.

\section{Probability of being surrounded by $m$ loops}
We have succeeded in guessing an exact formula for the probability $P\haak{L,m}$ that a face of the lattice on a cylinder of circumference $L$ is surrounded by $m$ loops.
Writing $P\haak{L,m}=Q\haak{L,m}/A_{\text{HT}}\haak{L}^{2}$, we conjecture that $Q\haak{L,m}$ is an integer given by:
\begin{equation}\label{plm}
Q\haak{L,m}=C_{L/2-m}\haak{L}+\sum_{r=1}^{L/4-m/2}(-1)^{r}C_{L/2-m-2r}\haak{L}\frac{m+2r}{m+r}\binom{m+r}{r}
\end{equation}
Here the $C_{p}\haak{L}$ are the absolute values of the coefficients of the characteristic polynomial of the $L\times L$ Pascal matrix:
\begin{equation}\label{pasc}
\det_{1\leq r,s\leq L}\rhaak{\binom{r+s-2}{r-1}-x\delta_{r,s}}=\sum_{n=0}^{L}C_{n}\haak{L}\haak{-x}^{n}
\end{equation}
The large $L$ behavior of the $C_{p}\haak{L}$ is not known well enough to be able to extract the  asymptotics
of $P\haak{L,m}$ from them. The form of the asymptotics of the
$P\haak{L,m}$ can be obtained, however, by applying Coulomb gas
techniques to the dense O$(n)$ loop model. On the basis of these
(non-rigorous) methods, explained in \cite{nienhuis, nienhuis2}, 
it is believed \cite{nijs} that $P\haak{L,0}$ behaves asymptotically as:
\begin{equation}\label{cgs}
P\haak{L,0}=L^{-5/48}\sum_{k=0}^{\infty}a_{k}L^{-k/2}
\end{equation}
In this sum $a_{1}=a_{2}=a_{5}=a_{6}=0$, consistent with the
renormalization group prediction.

It is possible to write $Q\haak{L,0}$ as a binomial determinant as follows. Observe that according to~\eqref{plm}:
\begin{equation}\label{plm=0}
Q\haak{L,0}=C_{L/2}\haak{L}+2\sum_{r=1}^{L/4}(-1)^{r}C_{L/2-2r}\haak{L}
\end{equation}

The relation $C_{L/2+p}\haak{L}=C_{L/2-p}\haak{L}$ (see \cite{lunnon}) allows one to rewrite this as
\begin{equation}\label{pl0det}
Q\haak{L,0}=\sum_{k=0}^{L}C_{k}\haak{L}\imath^{k-L/2}=\imath^{-L/2}\det_{1\leq r,s\leq L}\rhaak{\binom{r+s-2}{r-1}+\imath\delta_{r,s}}
\end{equation}
It thus follows from \eqref{cgs} that
\begin{equation}\label{reslt}
\det_{1\leq r,s\leq L}\rhaak{\binom{r+s-2}{r-1}+\imath\delta_{r,s}}=\imath^{L/2}L^{-5/48}A_{\text{HT}}\haak{L}^2
\sum_{k=0}^{\infty}a_{k}L^{-k/2}
\end{equation}
The first few nonzero coefficients in this expansion are approximately:
$$a_{0}\approx 0.81099753, \quad
a_{3}\approx  -0.028861, \quad
a_{4}\approx  0.021012.$$
The determinant in \eqref{reslt} can be interpreted in terms of cyclically symmetric plane partitions. A plane partition of an integer $N$ is an array of integers $n_{j,k}$, such that
$n_{j,k}\geq n_{j+1,k}$, $n_{j,k}\geq n_{j,k+1}$ and
$N=\sum_{j=1}^{\infty}\sum_{k=1}^{\infty}n_{j,k}.$
Plane partitions can be represented as a pile of unit cubes by introducing x, y, z-coordinates in $\mathbb{Z}^3$, and placing at position $\haak{j,k,0}$ a stack of $n_{j,k}$ cubes. A cyclically symmetric plane partition is a plane partition whose representation as a pile of cubes is invariant under a cyclic permutation of the x, y, z coordinates. In \cite{lozenge} it is shown that determinants of the form
\begin{equation}\label{omega}
\det_{1\leq r,s\leq L}\rhaak{\binom{r+s-2}{r-1}+\omega\delta_{r,s}}
\end{equation}
give a weighted enumeration of cyclically symmetric plane partitions in an $L\times L\times L$ box.
The weight assigned to a plane partition
is $\omega^{n}$, where $n$ is the number of unit cubes on the main
diagonal. In \cite{lozenge} closed form expressions of this determinant
are given and proved for the cases $\omega^6=1$. 

We close by using the determinant evaluation for
$\omega=\exp\haak{\imath\pi/3}$ to show that the conjectured expression
for $P\haak{L,m}$ is properly normalized. The conjecture \eqref{plm}
can be written as: 
\begin{equation}\label{normq}
\begin{split}
\sum_{m=0}^{L/2}Q\haak{L,m}&=\sum_{k=0}^{L/2}B_{k}C_{L/2-k}\haak{L}
\text{\quad
where $B_{0}=1$ and \quad}\\
B_{n}&=\sum_{r=0}^{n/2}(-1)^{r}\frac{n}{n-r}\binom{n-r}{r}
\end{split}
\end{equation}
for $n\geq 1$. The summand of (\ref{normq}), which we name $f\haak{n,r}$, 
satisfies the recurrence:
\begin{equation}
f\haak{n+2,r+1}-f\haak{n+1,r+1}+f\haak{n,r}=0
\end{equation}

Summing this over $r$ from $0$ to $n/2$ yields the following recurrence:
$B_{n+2}-B_{n+1}+B_{n}=0$.
From~\eqref{normq} it follows that $B_{1}=1$ and $B_{2}=-1$. This  yields
$B_{n}=2\cos\haak{\pi n/3}$.
Inserting this in \eqref{normq} and using the relation $C_{L/2+p}\haak{L}=C_{L/2-p}\haak{L}$ \cite{lunnon} yields:
\begin{equation}\label{normq2}
\sum_{m=0}^{L/2}Q\haak{L,m}=\exp\haak{-\imath\pi L/6}\det_{1\leq r,s \leq L}\rhaak{\binom{r+s-2}{r-1}+\exp\haak{\imath\pi/3}\delta_{r,s}}
\end{equation}
The r.h.s.\ of \eqref{normq2} should equal
$A_{\text{HT}}\haak{L}^2$. Using theorem 13 of \cite{lozenge} we
verified that this is indeed the case.

\section{Conclusion}
We have introduced so-called osculating walkers, defined as nonintersecting random walkers that at each vertex can make a left or right turn. These two possibilities are equally likely, unless the vertex has been visited before, in which case the
walker leaves the vertex via the unused edge. We have obtained several conjectures about such random walks on
a cylinder. Also we have obtained a conjecture for the asymptotics of the determinant $\det_{1\leq r,s\leq L}\rhaak{\binom{r+s-2}{r-1}+\imath\delta_{r,s}}$.

\end{document}